\documentstyle[aps,prl,preprint,floats,epsfig]{revtex}

\begin{document}

\preprint{\tighten\vbox{\hbox{\hfil CLNS 99/1634}
                        \hbox{\hfil CLEO 99-12}
}}

\title{Observation of Radiative Leptonic Decay of the Tau Lepton}  

\author{CLEO Collaboration}
\date{\today}

\maketitle
\tighten

\begin{abstract} 
Using 4.68 fb$^{-1}$ of $e^+e^-$ annihilation data collected with 
the CLEO II detector at the Cornell Electron Storage Ring (CESR) 
we have studied $\tau$ radiative decays 
$\tau^- \rightarrow {\nu}_\tau \mu^- \overline{\nu}_\mu \gamma$ and 
$\tau^- \rightarrow {\nu}_\tau {\it e}^- \overline{\nu}_{\it e} \gamma$.
For a 10 MeV minimum photon energy in the $\tau$ rest frame, the branching 
fraction for radiative $\tau$ decay to a muon or electron is measured to be 
$(3.61\pm0.16\pm0.35)\times10^{-3}$ or $(1.75\pm0.06\pm0.17)\times10^{-2}$,
respectively. The branching fractions are in agreement with Standard 
Model theoretical predictions.
\end{abstract}
\pacs{PACS numbers: 13.10.+q, 13.35.Dx, 14.60.Fg}
 
\newpage
{
\renewcommand{\thefootnote}{\fnsymbol{footnote}}

\begin{center}
T.~Bergfeld,$^{1}$ B.~I.~Eisenstein,$^{1}$ J.~Ernst,$^{1}$
G.~E.~Gladding,$^{1}$ G.~D.~Gollin,$^{1}$ R.~M.~Hans,$^{1}$
E.~Johnson,$^{1}$ I.~Karliner,$^{1}$ M.~A.~Marsh,$^{1}$
M.~Palmer,$^{1}$
% ...
\end{center} 
\small
\begin{center}
$^{1}${University of Illinois, Urbana-Champaign, Illinois 61801}\\
% ...
\end{center}
\setcounter{footnote}{0}
}
\newpage
Unconventional models for $\tau$ decay could lead to behavior inconsistent 
with the Standard Model in radiative $\tau$ decay \cite{MLP}. 
In one model $\tau$ decay occurs not only through the known s-channel exchange 
of a W-boson, but also through the s-channel exchange of an unknown X boson. 
In another model, $\tau$ decay occurs only through the exchange of the 
W-boson but the $\tau-\nu_\tau-$W vertex has anomalous radiative properties. 
In both cases, the radiative decay behavior of the
$\tau$ should be altered with respect to the Standard Model expectation. 

The data sample used in this work 
was acquired from ${\sl e^+}{\sl e^-}$ collisions at a center-of-mass  energy of 
$E_{cm} = 2\times E_{beam}\ \approx$ 10.6 GeV with the CLEO II detector \cite{DETE}
at the Cornell Electron Storage Ring (CESR). The total integrated luminosity of 
the data  is 4.68 fb$^{-1}$, corresponding to $N_{\tau\tau} = 4.3\times10^6$ $\tau$ pairs.
We search for $\tau^-\rightarrow\nu_\tau\ell^-\overline{\nu}_\ell\gamma$ ($\ell={\it e}$
or $\mu$) using the observed two-charged-track $\tau$ pair final states with a photon in the
lepton hemisphere as defined by the plane perpendicular to the thrust 
axis \cite{THRUST}: ${\it e^+}$+$\mu^-\gamma$, $\mu^+$+${\it e^-}\gamma$, 
${\it h^+}$+$\mu^-\gamma$, ${\it h^+}\pi^0$+$\mu^-\gamma$, ${\it h^+}$+${\it e^-}\gamma$, and 
${\it h^+}\pi^0$+$e^-\gamma$, where ${\it h^+}$ is a charged pion or 
kaon.\footnote{Charge conjugate states are included in this analysis.} 
The $\tau^+$ decay products are used to tag the events.

We select events with exactly two oppositely charged tracks with scaled momentum,
$x_\pm = p_\pm/E_{beam}$, satisfying $x_\pm < 0.9$ and with the angle
between the two tracks greater than $90^\circ$. We require exactly one charged 
track in each hemisphere. To suppress beam-gas interactions, the distance of closest 
approach of each track to the interaction point must be within 0.5 cm transverse to 
the beam direction, and 5 cm along it. Hadronic background is suppressed by requiring 
the total invariant mass of particles in each hemisphere to be less than the $\tau$ 
mass. In computing the invariant mass, we assign the pion mass to the charged hadron.  
We require the two-track acollinearity in azimuth, $\xi=||\phi_+-\phi_-|-\pi|$ 
where $\phi_+(\phi_-)$ is the azimuthal angle of the positively (negatively) charged
track, to satisfy $0.05<\xi<1.5$. The scaled missing momentum transverse 
to the beam, $x_t=p_t/E_{beam}$, and the angle of the missing momentum with 
regard to the beam line, $\theta_{miss}$, must satisfy $x_t>0.1$ and 
$|\cos\theta_{miss}|<0.8$ for all non-${\it h^+}\pi^0$ tag modes; for the two 
${\it h^+}\pi^0$-tag modes, only $x_t>0.05$ is required. These criteria effectively 
reduce potential contamination from non-$\tau$ QED events. 

Photons are defined as clusters in the calorimeter with energy $E_\gamma > 50$ MeV for 
$|\cos\theta|<0.71$, or 100 MeV when $0.71<|\cos\theta|<0.95$ where 
$\theta$ is the polar angle with respect to the beam axis. They 
are further required to pass a lateral shower shape requirement, that is $99\%$ 
efficient for isolated photons. No charged particle track can point to within 8 cm of 
a crystal used in the energy cluster. In the signal lepton hemisphere, we require 
that there be only one photon, and this photon must be in the region 
$|\cos\theta|<0.71$. In the tag hemisphere, if the tag is a 
lepton, then at most one unused photon is allowed; otherwise, at most two unused 
photons are allowed. Photons from $\tau$ radiative leptonic decays tend to be 
almost collinear with the final state lepton direction, hence we require 
$\cos\theta_{\mu\gamma}>0.96$ in the case of muonic decay and
$\cos\theta_{{\it e}\gamma}>0.99$ in the case of electronic decay. 

Identified electrons are required to have scaled momenta $x_\pm>0.1$ and 
$|\cos\theta|<0.71$.  The ratio of energy deposited 
in the calorimeter to track momenta for electron candidates must satisfy 
$E_\pm/p_\pm>0.85$. The drift chamber specific-ionization $(dE/dx)$ for electron
candidates must be no lower than two standard deviations below that expected for an 
electron. To exclude events in which a photon hides in the track's calorimeter 
shower, the criteria further require $E_\pm/p_\pm<1.1$. Muon criteria demand
that the track has $|\cos\theta|<0.71$ and deposit $E_\pm<0.3$ GeV in the 
calorimeter, consistent with a minimum-ionizing particle, and that there be hits in the 
muon detection system matched to the projected trajectory of the track. 
A muon candidate must also penetrate at least three hadronic interaction lengths 
for $p_\pm<2.0$ GeV/$c$ and five interaction lengths for $p_\pm>2.0$ GeV/$c$, 
corresponding to the first and second superlayers of the muon chambers. 
The tag $h$ is operationally defined as a charged track not identified as a 
lepton, with $p_\pm> 0.5$ GeV/$c$ and $|\cos\theta|<0.90$.
The $h\pi^0$ tag is defined as a reconstructed $\pi^0$ plus a 
charged track not identified as a lepton, and the charged track must satisfy  
$p_\pm>0.3$ GeV/$c$ and $|\cos\theta|<0.90$. A $\pi^0$ is reconstructed using two 
showers in the tag hemisphere that satisfy the photon criteria, 
except that only one of the showers is required to meet the lateral shower shape requirement. 
We require that the invariant mass of the two photons satisfy 
$120<m_{\gamma\gamma}<145$ MeV. We exclude events in which an  extra $\pi^0$ is found.

Additional criteria are applied to suppress mode-specific backgrounds. To reduce 
contamination from radiative QED processes 
${\it e^+}{\it e^-}\rightarrow{\it e^+}{\it e^-}\gamma$ 
and $\mu^+\mu^-\gamma$, the total energy of an event must satisfy $E_{tot}<7.5$ GeV for 
$h^++\mu\gamma$ and $h^++e\gamma$ modes. In the $h^++{\it e}\gamma$ mode, to further reduce 
background from ${\it e^+}{\it e^-}\rightarrow e^+e^-e^+e^-(\gamma)$ we require 
$E_{tot} > 2.8 $ GeV and the $h^+$ to satisfy $|\cos\theta|<0.71$.
In the three tag modes with the $\tau^-$ radiatively decaying to electron,
in order to reduce significant background from external bremsstrahlung, 
we require the distance of closest approach of the electron's
track to the interaction point to be within 0.08 cm transverse to the beam.  We further 
require the distance between the photon candidate shower and the electron shower in the 
calorimeter to be greater than 25 cm, in order to separate occasionally 
overlapping showers.

\begin{figure}[thb]
  \begin{center}
     {\resizebox{16.0cm}{16.8cm}{\includegraphics{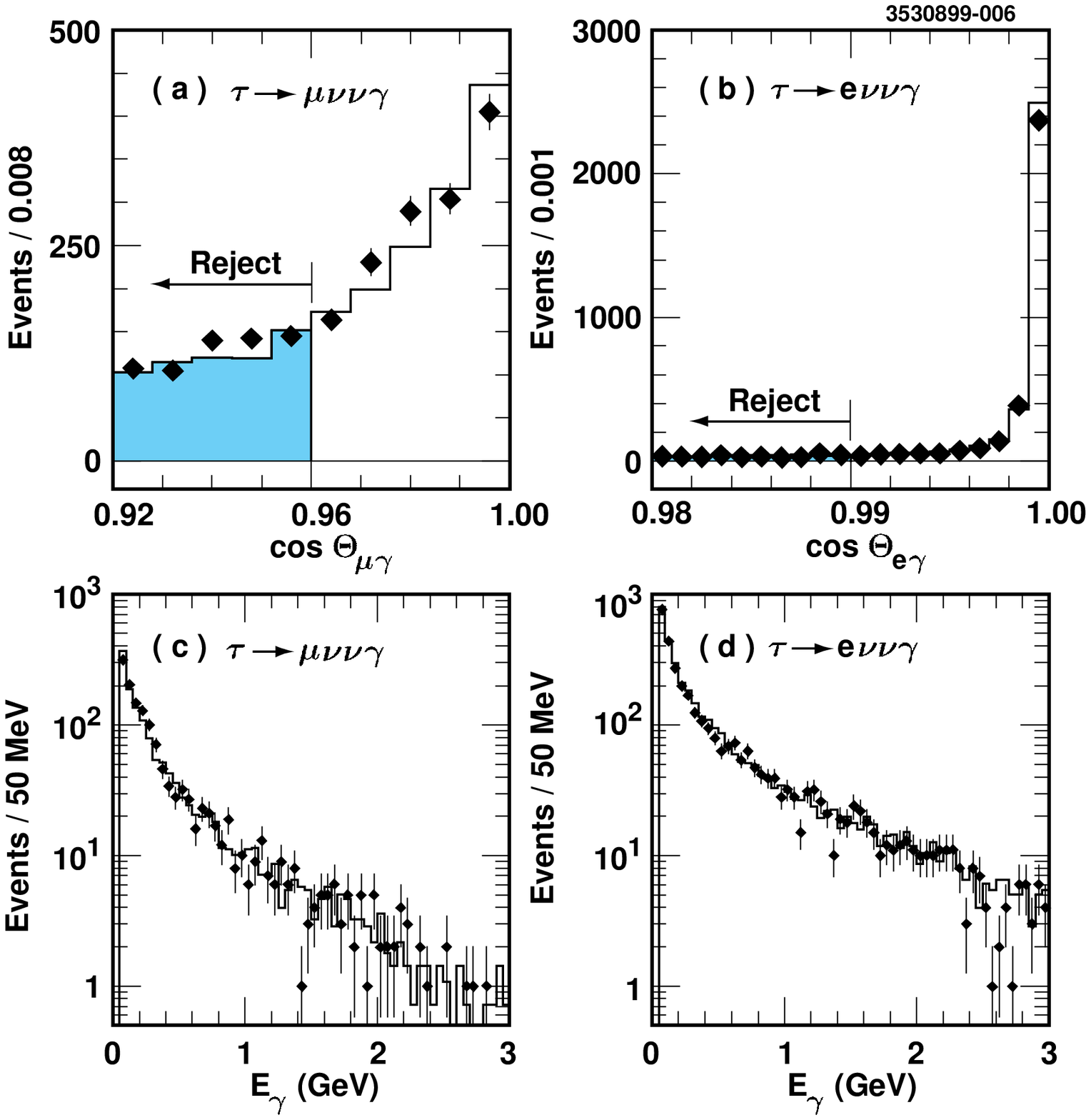}}}
    \vspace{0.8cm}
     \caption[]{ Distributions in $\cos\theta_{\ell\gamma}$ and $E_\gamma$
        for data (diamonds) and Monte Carlo (histogram) for both muonic and 
        electronic radiative decays of the $\tau$. Each distribution shown here is 
        the sum over all tag modes. Only events satisfying the $\cos\theta_{\ell\gamma}$ 
        requirement are used in the $E_\gamma$ distributions. }
      \label{fig:egcos}
  \end{center}
\end{figure}

The detection efficiencies and backgrounds are investigated with a Monte Carlo technique. 
We use the KORALB/TAUOLA \cite{SJZW} and PHOTOS \cite{EBBZ} MC packages to model the 
production and decay of $\tau$ pairs. The detector response is simulated using the 
GEANT program \cite{GRUN}. Generic Monte Carlo-produced $\tau$-pair decay events are 
used to study the kinematic distributions of the signal candidates and the backgrounds from 
$\tau$-pair decay sources. The $\cos\theta_{\ell\gamma}$ and $E_\gamma$ distributions 
from selected events for both muonic and electronic decays are shown in Fig. \ref{fig:egcos}. 
The figure shows that data and luminosity normalized Monte Carlo expectation agree well.
The small apparent disagreement at low photon energy is caused by a small relative 
inefficiency in the Monte Carlo reconstruction of low energy photons near muons, 
and is accounted for in the systematic error estimation. 
Using Monte Carlo-produced $\tau$-pairs in which one $\tau$ decays radiatively into
a lepton and neutrinos and the another $\tau$ decays generically, we determine the total 
detection efficiencies to be $(3.28\pm0.06)\%$ for radiative muonic decay  and  $(2.02\pm0.03)\%$ 
for radiative electronic decay.  
   
\begin{figure}[htb]
  \begin{center}
     {\resizebox{16.0cm}{9.5cm}{\includegraphics{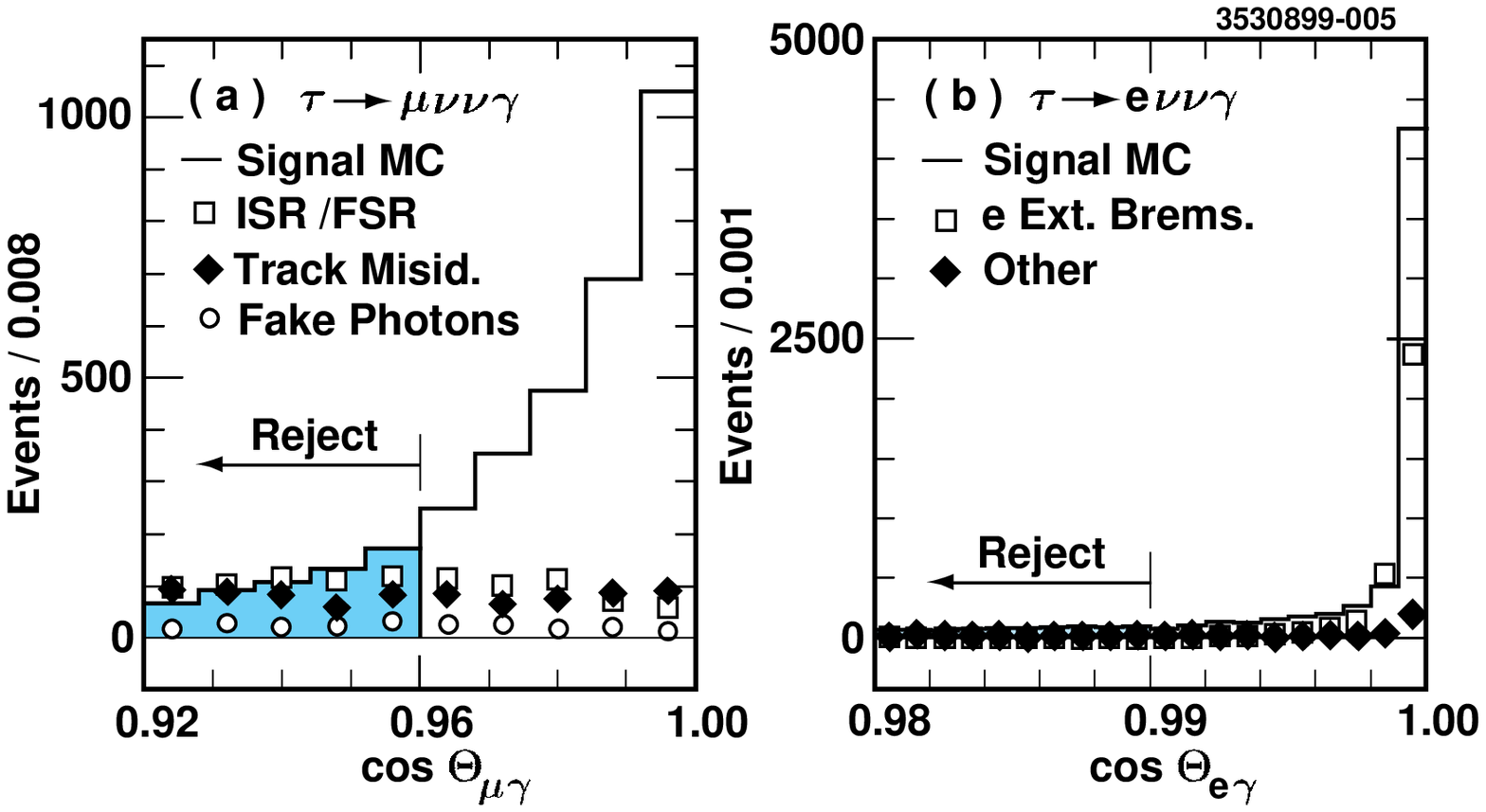}}}
     \vspace{0.8cm}
     \caption[]{ Distributions in $\cos\theta_{\ell\gamma}$ for signals and
        different tau source backgrounds from Monte Carlo simulations.} 
      \label{fig:c9298}
  \end{center}
\end{figure}

The backgrounds from $\tau$-pair decay sources relative to signals are shown 
with the $\cos\theta_{\ell\gamma}$ distributions in Fig. \ref{fig:c9298}. 
In the muonic decay case, the significant backgrounds
are ISR/FSR (initial state and final state radiation), track misidentification
(mostly other particles misidentified as a muon) and neutral showers faking 
photons. In the electronic decay case, the electron external bremsstrahlung
process is the only significant background; backgrounds such as other particles
misidentified as the electron are relatively small.
Figure \ref{fig:c9298} also shows that a photon from $\tau$ radiative decay to a lepton
tends to have a very small angle with respect to the final state lepton. Further from 
the lepton, background photons not related to the $\tau$ leptonic decay
completely dominate.

We investigate possible contamination from hadronic events by using the Lund 
simulation \cite{TSMB} and find that it is negligible. We rely upon Monte Carlo 
simulation of ${\it e^+}{\it e^-}$ to $\mu^+\mu^-(\gamma)$ \cite{RKVM}, 
${\it e^+}{\it e^-}(\gamma)$ \cite{BKFR,SJAD}, 
${\it e^+}{\it e^-}\mu^+\mu^-$ \cite{VMBU}, ${\it e^+}{\it e^-}\pi^+\pi^-$ \cite{VMBU}, 
and ${\it e^+}{\it e^-}{\it e^+}{\it e^-}$ \cite{JVER} final states 
to model backgrounds from these processes. All these background sources are 
small except in the two $h$ tag modes. In the case of muonic radiative decay 
with $h$ tag, we find that the two photon process 
${\it e^+}{\it e^-}\rightarrow{\it e^+}{\it e^-}\mu^+\mu^-$ contributes
$0.69\%$ to the selected sample in data and the QED process 
${\it e^+}{\it e^-}\rightarrow\mu^+\mu^-(\gamma)$ 
contributes another $0.46\%$. In the electronic decay case, the Monte Carlo predicts 
$0.42\%$ from the two-photon process 
${\it e^+}{\it e^-}\rightarrow{\it e^+}{\it e^-}{\it e^+}{\it e^-}$ and 
$0.29\%$ from the QED process 
${\it e^+}{\it e^-}\rightarrow{\it e^+}{\it e^-}(\gamma)$. As these processes are 
significantly suppressed by the selection criteria and their accurate normalization 
is difficult to verify, a total relative error of $100\%$
will be assigned in the final systematic errors. 
 
Branching fractions ${\cal B}(\tau^- \rightarrow \nu_\tau \mu^-\overline{\nu}_\mu\gamma)$
and ${\cal B}(\tau^-\rightarrow\nu_\tau{\sl e}^-\overline{\nu}_{\sl e}\gamma)$
are calculated for $E_\gamma>50$ MeV in the laboratory frame for each mode
and then converted into the $\tau$ rest frame for $E^{*}_{\gamma}>10$ MeV
by applying a boost factor assuming the Standard Model photon spectrum. 
The factor $\epsilon_{boost}$ is determined from
Monte Carlo simulation to be $0.754\pm0.007$ for muonic radiative decay and 
$0.762\pm0.003$ for electronic radiative decay. The branching fractions 
from three different tags are combined using a weighted average.
The measured branching fractions from data are compared with the  
theoretical predictions from Monte Carlo simulation.
Table \ref{table:relbr} summarizes the relative results
and Table \ref{table:abbr} shows the measured absolute branching fractions.
\begin{table}[htb]
 \vspace{0.2cm}
  \caption[]{ Branching fractions for $\tau^- \rightarrow \nu_\tau \mu^-\overline{\nu}_\mu\gamma$
              and $\tau^- \rightarrow \nu_\tau {\it e}^-\overline{\nu}_{\it e}\gamma$ relative to
              Standard Model Monte Carlo expectation for all tag modes and combined
              results for $E^{*}_{\gamma}>10$ MeV. Errors are statistical only.}
  \begin{center}
  \begin{tabular}{lccccc} 
  \ \ \ \ \  & ${\it e}$ tag & $\mu$ tag\ \ \ \ \  & ${\it h}$ tag\ \ \ \ \   
             & ${\it h}\pi^0$ tag\ \ \ \ \  & Total \ \ \ \ \   \\ \hline
   $\nu_\tau \mu^-\overline{\nu}_\mu\gamma$    
                        & $1.00\pm0.08$\ \ \ \ \  &   
                        & $0.98\pm0.07$\ \ \ \ \  & $0.96\pm0.07$\ \ \ \ \  
                        & $0.98\pm0.04$\ \ \ \ \  \\ 
   $\nu_\tau {\it e}^-\overline{\nu}_{\it e}\gamma$
                        &                         & $0.95\pm0.06$\ \ \ \ \  
                        & $0.90\pm0.06$\ \ \ \ \  & $0.97\pm0.05$\ \ \ \ \  
                        & $0.94\pm0.03$\ \ \ \ \  \\ 
 \end{tabular}
 \label{table:relbr}
\end{center}
\end{table}
\begin{table}[bht]
 \vspace{0.2cm}
  \caption[]{ Measured branching fractions 
              ${\cal B}(\tau^- \rightarrow \nu_\tau \mu^-\overline{\nu}_\mu\gamma)$ and
              ${\cal B}(\tau^- \rightarrow \nu_\tau {\it e}^-\overline{\nu}_{\it e}\gamma)$
              for $E^{*}_{\gamma}>10$ MeV and theoretical predictions from the Monte Carlo
              simulation. For data, the first error is statistical and the second
              one is systematic. For Monte Carlo, the error is based on the number of events 
              generated. Also listed is the ratio of
              ${\cal B}(\tau^- \rightarrow \nu_\tau {\it e}^-\overline{\nu}_{\it e}\gamma)$
              to ${\cal B}(\tau^- \rightarrow \nu_\tau \mu^-\overline{\nu}_\mu\gamma)$,
              ${\cal B}_{e\gamma}/{\cal B}_{\mu\gamma}$.} 
  \begin{center}
  \begin{tabular}{lcc} 
           & data \ \ \ \   & MC \ \ \ \   \\ \hline
    ${\cal B}(\tau^- \rightarrow \nu_\tau \mu^-\overline{\nu}_\mu\gamma)\ (\times10^{-3})$ 
           &   $3.61\pm0.16\pm0.35$\ \ \     &  $3.68\pm0.02$\ \ \    \\ 
    ${\cal B}(\tau^- \rightarrow \nu_\tau {\it e}^-\overline{\nu}_{\it e}\gamma)\ (\times10^{-2})$
           &   $1.75\pm0.06\pm0.17$\ \ \     &  $1.86\pm0.01$\ \ \    \\     
    ${\cal B}_{e\gamma}/{\cal B}_{\mu\gamma}$
           &   $4.85\pm0.27\pm0.57$\ \ \     &  $5.05\pm0.04$\ \ \    \\     
 \end{tabular}
 \label{table:abbr}
\end{center}
\end{table}

Systematic error estimates for $\tau^- \rightarrow \nu_\tau \mu^-\overline{\nu}_\mu\gamma$
and  $\tau^- \rightarrow \nu_\tau {\it e}^-\overline{\nu}_{\it e}\gamma$
are shown in Table \ref{table:syserrem}. The errors in the table are relative to
the final branching fraction. For muonic radiative decay, we estimate the error 
from photon reconstruction by varying the photon selection criteria
and also from a separate study of  ${\it e^+}{\it e^-}\rightarrow\mu^+\mu^-\gamma$ events. 
The trigger efficiency systematic error is obtained by a comparison of different 
triggers in data and Monte Carlo. We evaluate the muon misidentification 
systematic error by allowing a variation of the hadron to muon misidentification rate 
of $15\%$ as estimated from a sample of tracks in $\tau^+\tau^-$ events 
in which one $\tau$ decays to a lepton and the other $\tau$ decays to
${\it h^+}\pi^0$. The energy deposition of hadrons faking muons is not well
modeled in the Monte Carlo; therefore, we vary the muon maximum energy 
requirement to obtain its associated error. The integrated luminosity of the data at CLEO is 
measured with a relative error of $1\%$; this results in a relative error of $1.4\%$ on the 
total number of $\tau$ pairs produced in data, assuming a theoretical error of $1\%$ for 
the $\tau$-pair production cross section \cite{SJZW}. The uncertainty for the track 
finding efficiency is estimated from a visual scan of 
${\it e^+}{\it e^-}\rightarrow{\it e^+}{\it e^-}$  
events selected using shower information only and a study of pion finding efficiency 
in $\tau^+\tau^-$ events in which one $\tau$ decays to a lepton and the other $\tau$ 
decays to $3\pi^\pm(\pi^0)$. Other errors are small and we estimate these errors by 
either using an independent sample or by varying related individual requirements. 

The largest background to the decay 
$\tau^- \rightarrow \nu_\tau {\it e}^-\overline{\nu}_{\it e}\gamma$ comes from electron 
external bremsstrahlung. This process contributes about $40\%$ of the observed $\gamma$'s. 
Its systematic error contribution is estimated from comparisons of data and
Monte Carlo simulation for accepted $e^+e^-\gamma$ events from 
${\it e^+}{\it e^-}\rightarrow{\it e^+}{\it e^-}{\it e^+}{\it e^-}(\gamma)$.
The comparison indicates that external bremsstrahlung events in our Monte Carlo 
simulation are $(11\pm7)\%$ more likely than in data. This result is also confirmed by 
comparing the number of photon conversion events from $\pi^0$ decays between data and 
Monte Carlo. We estimate a propagated branching fraction error of $6.9\%$ by allowing a 
variation of as much as $18\%$ for this background. The error from photon reconstruction 
is estimated by varying the photon selection criteria.  All remaining errors are
estimated as in the muonic case. In calculating the systematic 
error for the ratio ${\cal B}_{e\gamma}/{\cal B}_{\mu\gamma}$, errors from the trigger, the 
number of $\tau$ pairs and the track-finding efficiency cancel.

\begin{table}[hbt]
 \vspace{0.3cm}
  \caption[]{ Summary of systematic errors from different sources for $\tau$ muonic and electronic
              radiative decays. }
  \begin{center}
  \begin{tabular}{lcc} 
   Source                           &   $\tau^- \rightarrow \nu_\tau \mu^-\overline{\nu}_\mu\gamma$
                                    &   $\tau^- \rightarrow \nu_\tau {\it e}^-\overline{\nu}_{\it e}\gamma$  \\ \hline
   external bremsstrahlung          &        $\approx 0.0$ &       $6.9$         \\
   photon reconstruction            &        $5.9$         &       $4.6$         \\      
   trigger                          &        $5.0$         &       $5.0$         \\      
   track misidentification          &        $4.4$         &       $1.1$         \\ 
   muon shower energy requirement   &        $3.6$         &       $NA$            \\
   $N_{\tau\tau}$                   &        $1.4$         &       $1.4$         \\  
   track-finding efficiency         &        $1.0$         &       $1.0$         \\ 
   non $\tau$ sources               &        $0.9$         &       $0.1$         \\
   ISR/FSR                          &        $0.8$         &       $0.2$         \\ \hline
   Total                            &        $9.8\%$       &       $9.9\%$         \\
 \end{tabular}
 \label{table:syserrem}
\end{center}
\end{table}

There have been measurements of $\tau$ radiative muonic decay from MARK II \cite{YWU} and 
OPAL \cite{GALE}. The OPAL result is more recent and more precise. 
OPAL reports a measurement of the 
branching fraction ${\cal B}(\tau^-\rightarrow {\nu}_\tau\mu^-\overline{\nu}_\mu\gamma) = 
(3.0\pm0.4\pm0.5)\times 10^{-3}$ for $E_{\gamma}^{*}>20$ MeV. 
Converting our result for $E_{\gamma}^{*}>10$ 
MeV to a result for $E_{\gamma}^{*}>20$ MeV gives a measurement of 
${\cal B}(\tau^-\rightarrow {\nu}_\tau\mu^-\overline{\nu}_\mu\gamma) = 
(3.04\pm0.14\pm0.30)\times 10^{-3}$, 
which is in excellent agreement with the OPAL result but with an error 
smaller by a factor of two. 
CLEO has previously observed $\tau$ radiative electronic decay \cite{BKHE},
but this is the first direct measurement of the branching fraction.

As pointed out in Refs. \cite{ASHV,WERA}, Lorentz structure parameters in $\tau$ decay that 
are difficult to measure directly in non radiative decays can also be investigated in 
radiative $\tau$ decay. For example, the probability $Q^{\ell}_{R}$ of the $\tau$ 
decaying into a right-handed charged daughter lepton is given by
$Q^{\ell}_{R}=\frac{1}{2}(1-\xi')$ ($\xi'$ = 1 in the Standard Model). If we could 
extract the Michel type parameter $\xi'$ by measuring the partial $\tau$ radiative 
decay rate \cite{ASHV}, then $Q^{\ell}_{R}$ could be limited. However, the differential 
$\tau$ radiative decay rate is most sensitive to $\xi'$ for photons emitted in the direction 
opposite to the daughter lepton, an area dominated by photons from other sources.
This indicates that we are unable to set useful limits using the experimental method 
described here. 

In summary, we have performed the first measurement of 
${\cal B}(\tau^-\rightarrow{\nu}_\tau{\it e}^-\overline{\nu}_{\it e}\gamma)$
and an improved measurement of 
${\cal B}(\tau^- \rightarrow {\nu}_\tau \mu^- \overline{\nu}_\mu \gamma)$ using 
the CLEO detector at the CESR. Within the errors
of the measurements we find that the magnitude of the decay rates and the kinematic
distributions agree with expectations of conventional electromagnetic and weak
interaction theory. We also conclude that it is not currently possible to set useful limits on the 
parameters proposed in \cite{ASHV,WERA} using the experimental method described 
in this letter.

We gratefully acknowledge the effort of the CESR staff in providing us with
excellent luminosity and running conditions. This work was supported by 
the National Science Foundation,
the U.S. Department of Energy,
the Research Corporation,
the Natural Sciences and Engineering Research Council of Canada, 
the A.P. Sloan Foundation, 
the Swiss National Science Foundation, 
and the Alexander von Humboldt Stiftung.

\end{document}